\documentclass[a4paper,twocolumn,floatfix]{revtex4}
\usepackage{graphicx}

\begin{document}

\title{Inherent size constraints on prokaryote gene networks due to
``accelerating" growth}

\author{M. J. Gagen and J. S. Mattick}

\affiliation{ARC Special Research Centre for Functional and
Applied Genomics, Institute for Molecular Bioscience, University
of Queensland, Brisbane, Qld 4072, Australia}

\email{m.gagen@imb.uq.edu.au}

\date{\today}

\begin{abstract}
Networks exhibiting ``accelerating" growth have total link numbers
growing faster than linearly with network size and can exhibit
transitions from stationary to nonstationary statistics and from
random to scale-free to regular statistics at particular critical
network sizes. However, if for any reason the network cannot
tolerate such gross structural changes then accelerating networks
are constrained to have sizes below some critical value.  This is
of interest as the regulatory gene networks of single celled
prokaryotes are characterized by an accelerating quadratic growth
and are size constrained to be less than about 10,000 genes
encoded in DNA sequence of less than about 10 megabases. This
paper presents a probabilistic accelerating network model for
prokaryotic gene regulation which closely matches observed
statistics by employing two classes of network nodes (regulatory
and non-regulatory) and directed links whose inbound heads are
exponentially distributed over all nodes and whose outbound tails
are preferentially attached to regulatory nodes and described by a
scale free distribution. This model explains the observed
quadratic growth in regulator number with gene number and predicts
an upper prokaryote size limit closely approximating the observed
value.
\end{abstract}

\maketitle

\section{Introduction}

The rapidly expanding field of network analysis, reviewed in
\cite{Dorogovtsev_02_10,Albert_02_47}, has provided examples of
networks exhibiting ``accelerating" network growth, where link
number grows faster than linearly with network size. For instance,
the Internet \cite{Faloutsos_99_25} appears to grow by adding
links more quickly than sites though the relative change over time
is small and the Internet appears to remain scale free and well
characterized by stationary statistics \cite{Vasquez_02_066130}.
Similarly, the number of links per substrate in the metabolic
networks of organisms appears to increase linearly with substrate
number \cite{Jeong_00_65}, the average number of links per
scientist in collaboration networks increases linearly over time
\cite{Dorogovtsev_00_33,Vasquez_00_0006132,Barabasi_01_0104162,Barabasi_02_590,Vasquez_03_056104},
and languages appear to evolve via accelerated growth
\cite{Dorogovtsev_01_2603}.

In the main, the chief focus of these studies has been on locating
parameter regimes allowing accelerating networks to maintain scale
free statistics and thereby to allow continued unconstrained
growth.  For example, an early study considered a growing network
receiving $N^\alpha$ new links for $\alpha>0$ when the network
size is at $N$ nodes, but restricted analysis to the  case
$\alpha\leq 1$ as ``Obviously, $\alpha$ cannot exceed 1 (the total
number of links has to be smaller than $N^2/2$ since one may
forbid multiple links)." and ``The density of connections in real
networks remains rather low all the time, so one may reasonably
assume that $\alpha$ is small." \cite{Dorogovtsev_01_025101}.
Equivalent limits were considered in Ref. \cite{Sen_0310513}.  In
such restricted parameter regimes networks could maintain scale
free statistics, though this result carries the implicit but
unexamined finding that alternate parameter regimes permit
transitions from stationary to nonstationary statistics. This
paper builds on these implicit findings.

Accelerating networks are more prevalent and important in society
and in biology than is commonly realized---see the survey in Ref.
\cite{Gagen_03_accel_survey}. In fact, any network that requires
functional integration and organization (where the activity of any
given node is dependent on the state of the network or different
subnetworks) is by definition an accelerating network, that is, as
the network expands, the proportion of the network devoted to
control and regulation expands disproportionately.  This in turn
means that all such networks, sooner or later, must be limited in
their size and complexity, which limitations can only be breached
by changing either the physical nature of the control architecture
(a state transition) or by reducing the functional integration. In
the latter case, where networks are hitting a complexity limit,
further growth in network size will likely display structural
transitions from randomly connected, to scale free statistics, to
densely connected and perhaps finally to fully connected
statistics. Should such networks be unable to successfully
complete these transitions for any reason, then it is likely that
network growth must cease entirely or that either a transition to
a nonaccelerating structure is required to permit further growth
or novel technologies must appear allowing the continuation of
accelerated growth. Exemplar accelerating networks displaying such
size limits or structural transitions include (a) all forms of
economic markets where the latest price offered by any participant
instantly affects all other participants, (b) industrial companies
and sectors implementing a Just-In-Time business model where any
worker can halt the entire production system, (c) error
propagation networks linking an error source with all affected
nodes as studied in software analysis and in models of the
propagation of diseases, bushfires, cracks, and electricity grid
failures, (d) in any dynamical system dependent on relative
quantities so changes in one node instantly affects every other
node such as relative transcription factor binding probabilities
or relative evolutionary fitness, (e) in computer hardware and in
cluster and grid supercomputer networks, and (f) in organizational
networks \cite{Gagen_03_accel_survey}. In fact, it is well
understood that social networks only take on small world
statistics when the network is large enough---in small towns
everyone one knows everyone else so social networks are
accelerating, and social networks make a transition to small world
statistics only as individual nodes saturate their connectivity
limits \cite{Watts_99_493}. Similar observations can be made about
the scale free Internet and World Wide Web---when sufficiently
small, these networks were likely accelerating until connectivity
capacities saturated forcing a transition to scale free structures
to permit further growth \cite{Gagen_03_accel_survey}.

This paper develops an accelerating network model of prokaryotic
(single celled) gene regulatory networks to investigate size and
complexity limits inherent in the adoption of an accelerating
architecture.  Because our focus is on structural transitions, we
explicitly do not need to restrict the degree of acceleration to
low values of $\alpha\approx 0$.  Rather, we permit this parameter
to take on any value including $\alpha>1$ and ensure that the
network is not saturated by making link formation probabilistic.
The resulting novel ``probabilistic" accelerating networks grow by
adding on average $pN^{\alpha}$ new links with $\alpha>0$ and
otherwise arbitrary provided the probability of adding a link is
suitably constrained $p\ll 1$ so that total link number remains
less than of order $N^2$.

The gene regulatory model presented here is motivated by
comparative genomics findings that the total number of regulatory
proteins controlling gene expression (links) scales quadratically
with the number of genes or operons (nodes) in prokaryotes
\cite{vanNimwegen_03_479,Croft_03_unpub}. This quadratic growth
results as the number of links made by a regulator exploiting
homology dependent (sequence specific) interactions scales
proportionally to the number of randomly drifting promotor
sequences or effectively, with gene number \cite{Croft_03_unpub}.
Hence, gene regulatory networks are inherently accelerating---the
probable number of links per regulator $pN^{\alpha}$ increases
linearly with node number with $\alpha=1$, so consequently, the
total number of links scales quadratically as $pN^{\alpha+1}$.  In
small and sparsely connected networks, most links come from
different regulators suggesting that regulator number also scales
quadratically with gene number, $pN^{\alpha+1}$. Such an
accelerating network would be characterized initially by sparse
connectivity at low gene numbers and subsequently by denser
connectivity at high gene numbers as networks attempt a transition
to a densely connected regime. If the evolving networks can
successfully make this transition, the evolutionary record will
display a transition in network statistics for some critical
network size $N_c$. Conversely, if these networks, optimized by
evolution in the sparse regime, are unable to make the transition
to the densely connected regime, the evolutionary record would
show a strict size limit $N\leq N_c$ at some critical network
size. But this is exactly what is observed. All prokaryotic gene
numbers and genomes are indeed of restricted size (less than about
10,000 genes with genomes of between 0.5 and 10 megabases
\cite{Casjens_98_33}), in contrast to the genomes of multicellular
eukaryotes (with for humans, about 30,000 genes and a genome of
about 3 gigabases \cite{Int_Human_genome_01_86,Venter_01_13}).
Ref. \cite{Croft_03_unpub} predicted the size limit $N_c\leq
20,000$ genes as continued genome growth requires the number of
new regulators to exceed the number of nonregulatory nodes.

A satisfactory model of prokaryotic gene regulatory networks
requires some novel features.  As mentioned above, we introduce
probabilistic link formation to allow rapid accelerated growth and
correspondingly stricter size limits. (A different but related
mechanism was introduced in Refs.
\cite{Liu_02_036112,Goh_02_108701} which considered the effects of
stochastic fluctuations in the number of added links with each
additional node.) In addition, we employ directed links and
partition nodes into two classes where ``regulators" can source
outbound regulatory links to regulate other nodes (both regulators
and non-regulators), while ``non-regulators" cannot source
outbound links. (Ref. \cite{Cheng_02_066115} has previously
considered networks of distinguishable nodes.) Further,
experimental evidence presented below indicates that the
distribution of inbound links is compact and exponential while the
distribution of outbound links is long-tailed and likely
scale-free.  As a result, the heads and tails of our directed
links are placed according to two distinct distributions.
Altogether, these features allow the reproduction of the observed
features of prokaryote gene regulatory networks and satisfactorily
predicts the maximum prokaryotic gene count.

Our approach reproducing accelerating network statistics for
growing prokaryote genomes complements and informs alternate
networking approaches seeking to deduce or simulate the regulatory
networks of particular organisms from gene perturbation and
microarray experiments
\cite{Wagner_01_1183,Bhan_02_1486,Lukashin_03_1909}.

In Section \ref{sect_overview_prok_networks} we canvass the
available literature to characterize the statistics of prokaryote
gene regulatory networks.  This then allows the construction of
accelerating growth network models in Section
\ref{sect_accelerating_models} where we use the continuous
approximation and simulations to analyze network statistics. The
size constraints inherent in accelerating prokaryote regulatory
networks are modelled in Section \ref{sect_size_constraints}.

\section{Overview of prokaryote gene networks}
\label{sect_overview_prok_networks}

Ongoing genome projects are now providing sufficient data to
usefully constrain analysis of the gene regulatory networks of the
simpler organisms. Ref. \cite{vanNimwegen_03_479} first noted the
essentially quadratic growth in the class of transcriptional
regulators ($R$) with the number of genes ($N_g$) in bacteria with
the observed results
\begin{equation}
   R\propto \left\{
     \begin{array}{ll}
       N_g^{1.87\pm0.13}, & \mbox{transcriptional regulation} \\
        & \\
       N_g^{2.07\pm0.21}, & \mbox{two component systems} \\
        & \\
       N_g^{2.03\pm0.13}, & \mbox{transcriptional regulation} \\
        & \\
       N_g^{2.16\pm0.26}, & \mbox{transcriptional regulation}. \\
     \end{array}
    \right.
\end{equation}
Here, the top two lines refer to different classes of regulators
while the bottom two lines are the results of a crosschecking
analysis of two alternate databases. Quoted intervals reflect 99\%
confidence limits \cite{vanNimwegen_03_479}. The explanation for
this quadratic growth was that each additional transcription
factor doubles the number of available dynamical states which, it
was posited, allows for a doubling in the fixation probabilities
for this class of genes.

\begin{figure}[htbp]
\centering
\includegraphics[width=0.9\columnwidth,clip]{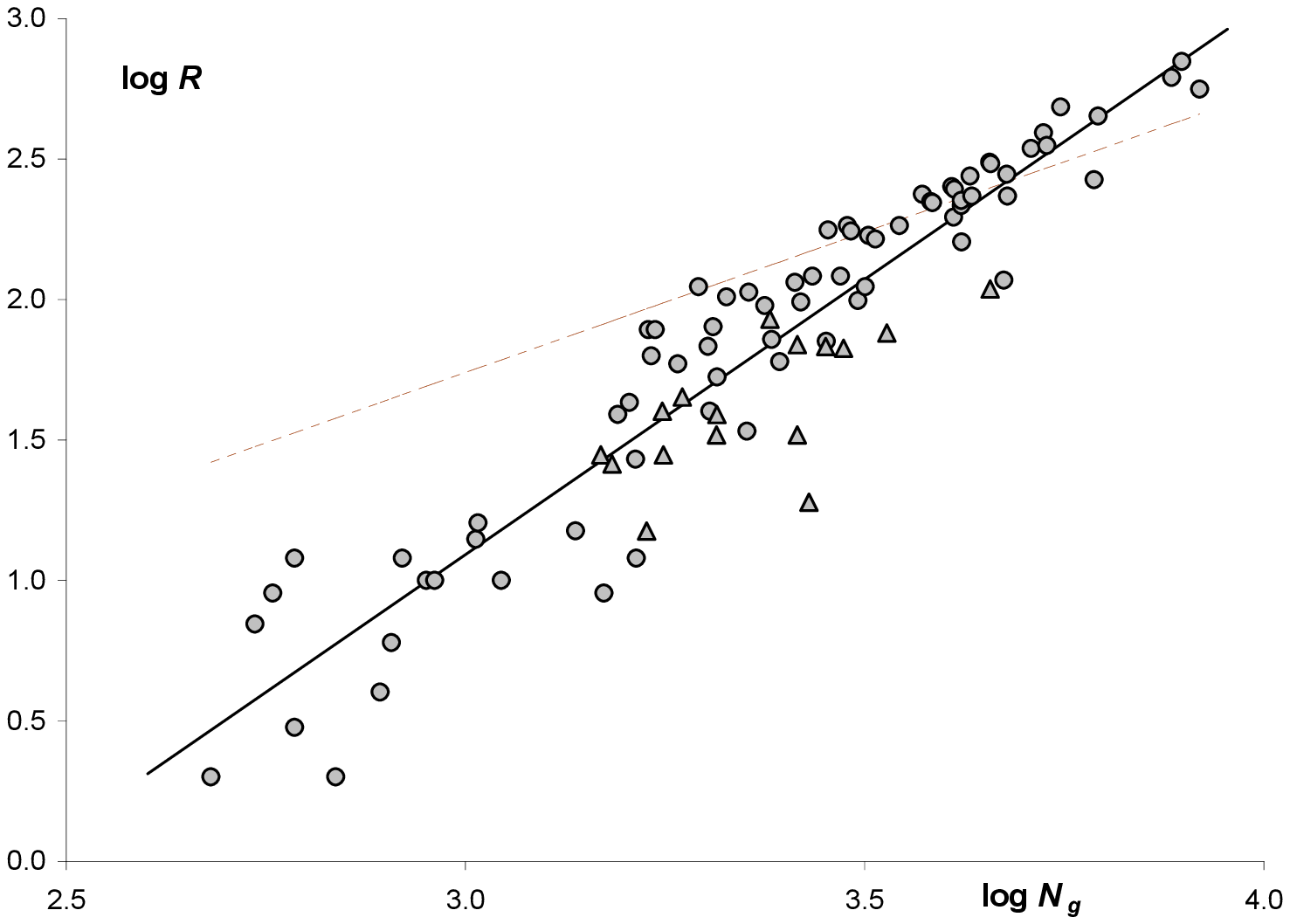}%
\hspace{-4cm}
\includegraphics[width=0.5\columnwidth]{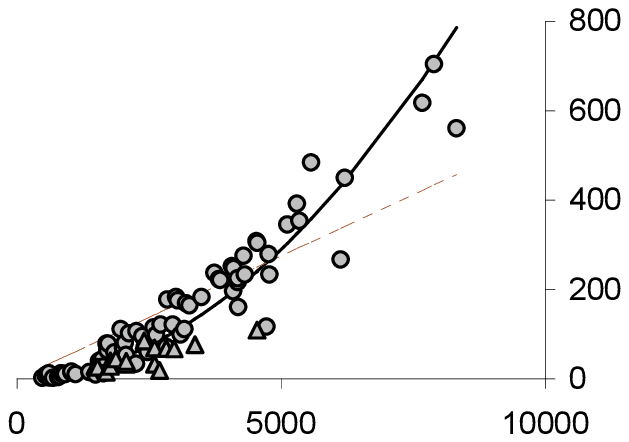}
\caption{\em Double-logarithmic plot of regulatory protein number
($R$) against total gene number ($N_g$) for bacteria (circles) and
archaea (triangles), adapted from Ref.
\protect\cite{Croft_03_unpub}.  The log-log distribution is well
described by a straight line with slope $1.96\pm 0.15$ ($r^2=
0.88$, 95\% confidence interval indicated), corresponding to a
quadratic relationship between regulator number and genome size.
The inset shows the same data before log-transformation
\protect\cite{Croft_03_unpub}. Dashed lines show the best linear
fit to the data $R=(0.055\pm0.004)N_g$ ($r^2=0.75$).}
\label{f_prok_reg_data}
\end{figure}

As noted above, Ref. \cite{Croft_03_unpub} provides an alternate
theoretical analysis predicting quadratic growth in any regulatory
network exploiting homology dependent interactions and analyzed 89
bacterial and archeael genomes to determine the relations
\begin{equation}     \label{eq_Croft_results}
   R= \left\{
     \begin{array}{ll}
       aN_g^b=(1.6\pm0.8)10^{-5}N_g^{1.96\pm0.15} & (r^2=0.88) \\
        &  \\
       pN_g^2=(1.10\pm0.06)10^{-5}N_g^{2} & (r^2=0.87) \\
        &  \\
       cN_g=(0.055\pm0.004)N_g  & (r^2=0.75). \\
     \end{array}
    \right.
\end{equation}
In this paper, accelerating networks will be based on the
quadratic second line (while nonaccelerating models presented in
later work will work with the linear third line
\cite{Gagen_0312022}). In all cases, the limits reflect 95\%
confidence levels. For completeness, the data is shown in Fig.
\ref{f_prok_reg_data}. The observed quadratic growth implies an
ever growing regulatory overhead so there will eventually come a
point where continued genome growth requires the number of new
regulators to exceed the number of nonregulatory nodes, and based
on this, Ref. \cite{Croft_03_unpub} predicted an upper size limit
of about 20,000 genes, within a factor of two of the observed
ceiling.

Earlier surveys of bacterial genomes noted that larger genomes
harboured more transcription factors per gene than smaller ones
\cite{Cases_03_248}, with this trend attributed to the need in
larger genomes for a more complex network of regulatory proteins
to achieve coordinated expression of a larger set of cellular
functions, and to selection in complex environments leading to
enrichment in transcription factors allowing regulation of gene
expression and signal integration. A similar upward trend in the
proportion of regulators as a fraction of genome size with
increasing genome size was observed in Ref. \cite{Stover_00_959}
attributed to a need for an increasing responsiveness in diverse
environments, with confirming observations in Ref.
\cite{Bentley_02_141}.

Prokaryotes typically group their DNA encoded genes in operons,
co-regulated functional modules of average size 1.70 genes each in
{\em E. coli} which value we treat as typical though in reality,
operon size decreases slightly with genome size
\cite{Cherry_03_40}. Each operon can be either unregulated and so
constitutively or stochastically expressed or subject to
combinatoric regulation by multiple regulatory protein
transcription factors binding to each operon's promotor sequence.

Again assuming that {\em E. coli} is typical, any given regulatory
protein affects an average of about 5 operons with this
distribution being long tailed \cite{Shen_Orr_02_64} so the
majority of regulators affect only one operon though some
regulators (CRP) can affect up to 71 operons or 133 genes
\cite{Thieffry_98_43}. (This latter reference estimated that each
regulator controls on average 3 genes.)  More recent estimates
have the transcription factor CRP, a global sensor of food levels
in the environment, regulating up to 197 genes directly and a
further 113 genes indirectly via 18 other transcription factors
\cite{MandanBabu_03_1234}.  (To observe the long tailed
distribution, see Fig. 2 of Ref. \cite{Thieffry_98_43} and Fig. 4
of Ref. \cite{MandanBabu_03_1234}.)

However, the number of inputs taken by an operon is characterized
by a compact exponential distribution with a rapidly decaying tail
so the majority of regulated operons are controlled by a single
regulator while very few regulated operons are controlled by four,
five, six or seven regulators
\cite{Thieffry_98_43,Shen_Orr_02_64,MandanBabu_03_1234}. In
particular, Ref. \cite{Thieffry_98_43} examined 500 regulatory
links from about 100 regulators to almost 300 operons to estimate
that each regulated operon takes on average 2 inputs though Fig. 2
of this reference suggests an average input number of about 1.5.
Similarly, Ref. \cite{Shen_Orr_02_64} suggests that 424 regulated
operons receive 577 links giving an average input number of 1.4,
while Ref. \cite{MandanBabu_03_1234} estimates that 327 regulated
operons receive 524 links giving an average input number of 1.6.

\section{Accelerating prokaryote network models}
\label{sect_accelerating_models}

We extend the gene network model of Ref. \cite{Thieffry_98_43} to
construct an accelerating network model of prokaryote regulatory
gene networks.  Prokaryotes typically pack their $N_g$ genes into
a lesser number of $N=N_g/g_o$ co-regulated operons where we
assume that operons contain exactly $g_o=1.70$ genes.    Of the
existing operons, $O_r$ are regulated operons and $O_u=N-O_r$ are
unregulated operons.  Of the total number of operons, there are
$R$ regulatory operons whose regulatory interactions are directed
links from regulatory operons to regulated operons. Under the
assumption that there is only one regulatory gene per regulatory
operon, the observed quadratic relation of Eq.
\ref{eq_Croft_results} becomes
\begin{equation}   \label{eq_reg_No}
    R=pg_o^2N^2.
\end{equation}
When regulators and regulatory links are very rare, i.e. when
genomes are small, it is likely that every new link is associated
with a new regulator so the number of links varies roughly
quadratically with operon number.  We write
\begin{equation}   \label{eq_link_No}
    L=lN^2,
\end{equation}
where $l$ denotes the probability of forming a particular
beneficial link per operon.  The value for $l$ will be
approximately $pg_o^2$, but the exact relation must be derived
from the details of the implemented model.

Each regulatory link between nodes is directed, and characterized
by two distinct distributions describing respectively the
placement of the heads and tails of each link.  Only a relatively
few nodes are regulatory, and of these, the number of outbound
link tails per regulatory node are described by a size dependent
long-tailed distribution with average about $\langle
t\rangle\approx 5$. Such a long-tailed distribution requires that
link tails be preferentially attached to an existing regulatory
operon or equivalently, the associated regulated operon must
possess one promotor binding site (among others) that binds that
particular regulator.  Consequently, the preferential selection of
regulators means that the promotor sequences of newly regulated
nodes cannot be randomly chosen---randomly drifting promotor
sequences would be as likely to match any one regulator as
another. A plausible physical explanation for the preferential
attachment of link tails to existing regulators is that newly
fixated operons come largely from gene duplication events
\cite{Yanai_00_2641} where some of the duplicated promotor binding
sites are under strong selective constraint while other binding
sites and the operon genes can drift freely. Gene duplication then
implies that in a genome of size $N$ operons, if some regulator
$n_j$ has $t_{jN}$ outbound regulatory links to approximately
$t_{jN}$ regulated operons, then the probability that a newly
fixated operon is also regulated by $n_j$ is simply the proportion
of such regulated operons in the genome, or $t_{jN}/N$. This
implements the required preferential attachment as the resulting
rate of growth in the number of links attached to node $n_j$ is
also then proportional to $t_{jN}$.  If there is also some
probability of the appearance of novel promotor sequences, these
combined processes suffice to produce the observed scale free
distributions. This model is roughly consistent with recent
estimates of the relative contributions to prokaryote genome
growth which suggest that horizontal gene transfer rates
$\gamma_h$ are roughly one third of gene loss rates
$\gamma_h=\gamma_l/3$ and roughly one half of vertical inheritance
or gene genesis rates $\gamma_h=\gamma_v/2$ leading to roughly
constant sized genomes over long times (as $\dot{N}\approx
\gamma_h+\gamma_v-\gamma_l\approx0$), while ``it is remarkable
that phylogenetic distributions of at least 60\% [and up to 75\%]
of protein families can be explained merely by vertical
inheritance." \cite{Kunin_03_1589}. Similarly, three quarters of
examined transcription factors in Ref. \cite{MandanBabu_03_1234}
were two-domain proteins with shared domain architectures leading
to the estimate that about 75\% of transcription factors have
arisen as a consequence of gene duplication (though the joint
duplication of regulatory regions and of regulated genes or of
transcription factors together with regulated genes is more rare).
A further implication of these gene duplication processes is that,
in the main, regulators can only appear on entry to the genome---a
potential regulator lacking any target matches in a given genome
will never form any links when most operons arise from promotor
preserving duplication events. This allows us to considerably
simplify our model, and hereafter, we only allow regulators to
appear on their entry to the genome. Of course, more realistic but
considerably more complicated models are possible.

In contrast to the relatively small number of regulatory nodes,
all nodes can themselves be regulated by inbound links and in
fact, can be multiply regulated as promotor regions can contain
more than one binding site.  Further, the many used and unused
promotor region binding sites broadly sample the space of possible
binding sites so only a small fraction of nodes will be regulated
by any one regulator. As a result, the number of inbound link
heads per node is described by a size dependent exponential
distribution with a low average of $\langle h\rangle\approx 1.5$
as typically results from the random or non-preferential
attachment of inbound links to operon promotor sequences.

\begin{figure}[htbp]
\centering
\includegraphics[width=\columnwidth,clip]{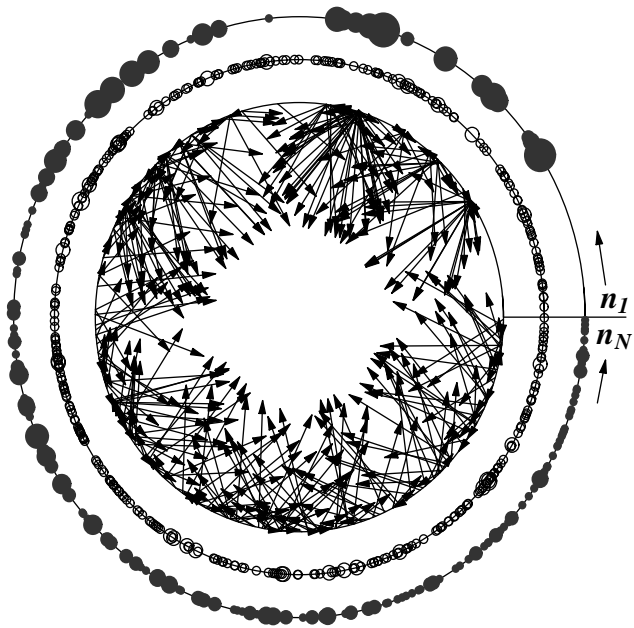}
\caption{\em An example statistically generated E. coli genome
using the later results of this paper where (for convenience only)
operon nodes numbered $n_1, \dots, n_N$ are placed sequentially
counterclockwise on a circle in their historical order of entry
into the genome. The filled points on the outer circle locate
regulators and have radius indicating the number of outbound
regulatory links. The open points on the middle circle locate
regulated operons and have radius indicating the number of inbound
regulatory inputs.  The arrows in the inner circle show all
directed regulatory links. }\label{f_e_coli_network}
\end{figure}

We suppose that the operon network grows by the sequential
addition of numbered nodes $n_k$ for $1\leq k\leq N$, and that at
network size $k$, node $n_i$ ($1\leq i\leq k$) has $t_{ik}$
outbound tails and $h_{ik}$ inbound heads.  We do not model the
many trials of potential genes over many generations and merely
include fixated genes in our count---that is, drifting sequence is
not counted as part of the fixated genome.  This further implies
the sequence of established nodes is under severe selective
constraint and unable to drift so consequently new links cannot be
added between existing nodes. (If a proportion $fN$ of existing
nodes can explore novel sequence space in time $dt$, then the
number of new regulators increases as $dR\propto fN^2dt$, and as
$N$ is itself a function of time, this integrates to generate a
non-quadratic relation between regulator and operon number which
is not observed.)

For clarity, Fig. \ref{f_e_coli_network} preempts later
calculations and depicts a statistically generated version of an
{\em E. coli} genome where nodes are placed sequentially
counterclockwise in a circle (for convenience only).  Alternative
genome models may be distinguished by the age distribution of
regulators, regulated operons and their link numbers, and these
are indicated in this figure. In particular, Fig.
\ref{f_e_coli_network} shows a highly nonuniform distribution of
regulators and outbound link numbers with gene age in contrast to
a uniform distribution of regulated operons and of inbound link
numbers. (It will turn out that these latter age-independent
distribution are only present when regulator number grows
quadratically with genome size.)

These distributions result from the physical processes underlying
the formation of regulatory links in prokaryotes. As discussed
above, a substantial proportion of the gene regulation network of
prokaryotes is enacted via homology dependent interactions as when
sequence specified protein transcription factors bind to specific
promoter sequences. The undirected nature of evolutionary searches
means that gene regulatory networks fundamentally exploit the same
sequence matching algorithms used in comparative genetics where
the probability of obtaining matches between a single given trial
sequence of some small fixed length and an entire genome scales
proportionately to genome length---doubling genome length doubles
the probability of a match. Hence, the expected number of links
formed per regulator scales linearly with present genome size. As
the number of source trial sequences also scales with genome
length, the expected number of matches between all regulators and
all regulated operons scales quadratically with genome length, or
effectively, with operon number assuming constant sized operons
over the evolutionary record.

As a consequence, on entry into the genome, each new gene has some
probability of being a regulator dependent firstly on its
suitability to bind DNA and secondly on the linearly increasing
expected number of acceptable binding targets present in the
genome on entry (or at later times).  As discussed above, the
predominance of vertical gene genesis events allows a simplified
model wherein the probability of a new node being regulatory is
determined solely by the number of available links present at the
time of entry. We assume then that on entry into the genome each
new node $n_k$ can form $2k-1$ links with nodes $n_1, \dots, n_k$
consisting of a single self-regulatory link from node $n_k$ to
itself with probability $l$, $(k-1)$ regulatory outbound links to
the existing nodes each with equal probability $l$, and, provided
that sufficient regulators already exist, $l(k-1)$ inbound
regulatory links from some subset of the existing regulators
chosen according to preferential attachment.  (For consistency, we
can only add $\approx lk$ distinct regulatory links to node $n_k$
provided there are at least this many regulators in existence.
From Eq. \ref{eq_reg_No}, the average number of regulators
$pg_o^2k^2$ must be greater than the number of regulatory links
$lk$, and this will be satisfied for $k>l/(pg_o^2)\approx 1$.)  As
a result, the total number of heads or tails attached to node
$n_k$ on entry to the genome ranges between $0$ and $k$, with each
link formed with probability $l$. Hence, the respective
probabilities that the initial number of heads $h_{kk}=j$ or the
initial number of tails $t_{kk}=j$ for node $n_k$ is
\begin{equation}   \label{eq_P_j_k_dist}
 P(j,k) = {k \choose j} l^j (1-l)^{k-j},
\end{equation}
with the proviso that all the inbound links can only be added to
node $n_k$ if there is a sufficient number of regulators among the
nodes $n_1, \dots, n_k$. The average number of inbound and
outbound links is identical, $\langle t_{kk}\rangle=\langle
h_{kk}\rangle=lk$ showing linear growth in link number with
increasing network size. The addition of node $n_k$ and its links
will increase the probable number of heads attached to earlier
nodes $n_j$ for $1\leq j\leq (k-1)$ so $h_{jk}\geq h_{jj}$, while
the probable number of tails outbound from node $n_j$ increases
$t_{jk}\geq t_{jj}$ if and only if that node is regulatory with
$t_{jj}>0$.

As regulators can only be created on entry to the genome, the
distribution of regulators at any time is specified by the
distribution $P(j,k)$ for $t_{kk}$.  Using Eq.
\ref{eq_P_j_k_dist}, the probability that node $n_k$ is a
regulator is $1-P(0,k)$, so for a network of $N$ nodes, the
predicted total number of regulators is
\begin{eqnarray}      \label{eq_reg_density}
  R &=&  \sum_{k=1}^{N} \left[ 1- (1-l)^k \right]  \nonumber  \\
   &=& N- \frac{1-l-(1-l)^{N+1}}{l}   \nonumber  \\
   &\approx & \frac{l}{2}N(N+1).
\end{eqnarray}
The exact top line shows the expected behaviour for the number of
regulators in the respective limits $l\rightarrow 0$ giving
$R\rightarrow 0$, and $l\rightarrow 1$ giving $R\rightarrow N$.
The approximate relation in the third line can be compared to the
observed Eq. \ref{eq_reg_No} and immediately suggests $l\approx
2pg_o^2$, while a fit to the more accurate top line gives the
connection probability as
\begin{equation}      \label{eq_l_assignment}
    l = 1.15 \times 2 p g_o^2 \;=\; 7.31 \times 10^{-5}.
\end{equation}
This probability value suggests an average promotor binding site
length of $-\log_4 l=6.9$ bases.  The average number of links per
regulator using the second line of Eq. \ref{eq_reg_density} is
then approximately $L/R\approx 2$, while the more accurate top
line with $N=2528$ operons for {\em E. coli} \cite{Cherry_03_40}
gives $L/R=2.12$, about a factor of two from the observed value of
$5$ for {\em E. coli} \cite{Shen_Orr_02_64}.

\subsection{Random distribution of regulated operons}

The distribution of link heads for all nodes (with possession of a
link head designating a regulated node), can be straightforwardly
calculated under the assumption that the $t_{kk}\approx lk$ new
tails added with node $n_k$ are randomly distributed across the
$k$ existing nodes so on average, each existing node receives $l$
links.  To build insight, it is useful to consider the general
case where $t_{kk}\approx h_{kk}\approx lk^\alpha$ for $\alpha\geq
0$. Setting $\alpha=0$ adds with some probability a constant
number of links with each new node, $\alpha=1$ adds a linearly
growing number of probable links with each new node, $\alpha=2$
adds a quadratically growing number of probable links with each
new node, and so on.  The total number of links present in the
network is then
\begin{equation}    \label{eq_total_links}
    \int_0^N 2lk^\alpha = \frac{2lN^{\alpha+1}}{\alpha+1}
\end{equation}
The continuous approximation
\cite{Barabasi_99_17,Barabasi_99_50,Dorogovtsev_01_056125} for
links randomly distributed over $k$ existing nodes determines the
number of inbound head links for node $n_j$ according to
\begin{eqnarray}    \label{eq_continuum_heads}
  \frac{\partial h_{jk}}{\partial k}&=& \frac{t_{kk}}{k} \nonumber \\
   & = & l k^{\alpha-1}.
\end{eqnarray}
This can be integrated with initial conditions $h_{jj}\approx
lj^\alpha$ at time $j$ and final conditions $t_{jN}\approx
lN^\alpha$ at time $N$ to give
\begin{equation}    \label{eq_hjn}
  h_{jN} = \left\{
  \begin{array}{cc}
    l + l \ln\frac{N}{j} & \mbox{if }\alpha=0 \\
      &  \\
    \frac{l}{\alpha}N^\alpha + \frac{l(\alpha-1)}{\alpha}j^\alpha & \mbox{if }\alpha>0. \\
  \end{array}
  \right.
\end{equation}
Integration of these link numbers over all node numbers $j$ gives
the required total number of links as in Eq. \ref{eq_total_links}.
For $0\leq\alpha<1$, the number of links per node is monotonically
decreasing with node number. However, for $\alpha=1$ and only in
this case, the final distribution is independent of node number
$j$ because earlier nodes receive exactly enough links from latter
nodes to balance the initially biased distribution of heads
$h_{jj}\approx lj$, so in the end, all nodes receive on average
the same number of inbound regulatory links $\langle
h_{jN}\rangle=lN$ for $1\leq j\leq N$. For faster acceleration
rates, $\alpha>1$, the number of links per node is monotonically
increasing as later nodes receive a greater number of links on
entry to the genome and this imbalance is not corrected.

The possibility of monotonically increasing numbers of links with
node number in accelerating networks has not previously been
considered. This possibility requires modifying the usual
continuum approach
\cite{Barabasi_99_17,Barabasi_99_50,Dorogovtsev_01_056125} so the
final inbound link distribution is obtained via
\begin{eqnarray}         \label{eq_final_link_dist}
  H(k,N) &=& \frac{1}{N} \int_0^N dj \; \delta(k-h_{jN})\nonumber \\
  & = & \pm \frac{1}{N} \left( \frac{\partial h_{jN}}{\partial j}
  \right)^{-1}  \mbox{at }[j=j(k,N)],
\end{eqnarray}
where $j(k,N)$ is the solution of the equation $k=h_{jN}$. The top
line is used when all nodes possess the same average link number
while the second line is applicable with the plus (negative) sign
when the average numbers of links per node is monotonically
increasing (decreasing) with node number. Non-monotonic cases
require alternate approaches.

\begin{figure}[htbp]
\centering
\includegraphics[width=\columnwidth,clip]{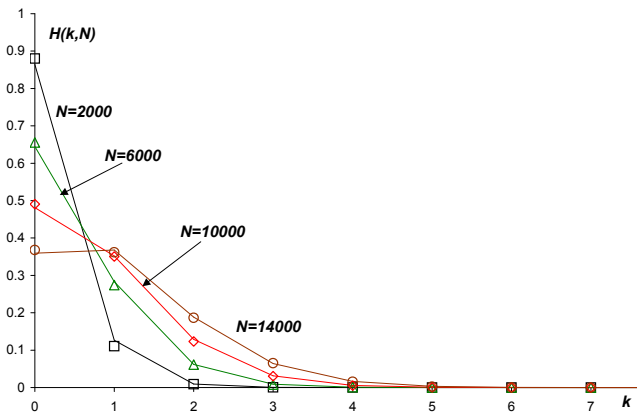}
\caption{\em  A comparison of the predicted distribution of
inbound link numbers per node $H(k,N)$ (solid lines) against that
observed in simulated networks of various sizes (indicated points)
with quadratic growth in the total probable number of randomly
attached links. \label{f_simulation_heads}}
\end{figure}

Under quadratic growth in total link number when $\alpha=1$, and
only in this case, the final distribution of link heads is
independent of node number and evaluated using Eq.
\ref{eq_final_link_dist} to give
\begin{eqnarray}
   H(k,N) &=& \frac{1}{N} \int_0^N dj \; \delta(k-lN)   \nonumber \\
   &=& \delta(k-lN).
\end{eqnarray}
As expected, a compact final link distribution results when all
nodes have an average of $t_{jN}=lN$ inbound regulatory links at
time $N$.  This distribution calculated under the continuous
approximation equates to one where in reality, each node receives
a controlling head with probability $l$ from every other node
(though in practise, the total number of received links is of
order unity). Hence, for any node in a network of size $N$, the
actual probability of having $k$ heads is
\begin{equation}    \label{eq_h_kNo_final}
     H(k,N) = {N \choose k} l^k (1-l)^{N-k}.
\end{equation}
A network simulation with linear growth of link numbers per node
model (Eq. \ref{eq_P_j_k_dist}) serves to validate this predicted
final distribution.  Fig. \ref{f_simulation_heads} compares the
predicted distribution $H(k,N)$ against observed distributions for
typical simulated networks of various sizes with negligible
discrepancies.

For a network of size $N$, the probability that any given operon
is unregulated is $H(0,N)$ so the expected number of unregulated
operons summed over all $N$ nodes is
\begin{equation}   \label{eq_unreg_operons}
    O_u= N(1-l)^N.
\end{equation}
This determines the number of regulated operons as
\begin{equation}   \label{eq_reg_operons}
    O_r=N-O_u=N\left[1- (1-l)^N\right],
\end{equation}
showing the expected behaviour as $l\rightarrow 1$ giving
$O_r\rightarrow N$ and $l\rightarrow 0$ giving $O_r\approx lN^2=L$
as each of the sparsely distributed links hits a distinct
regulated operon. We note that random gene duplication and
deletion events will not change the $H(k,N)$ distribution (other
than changing $N$) as all nodes are identically connected on
average. The $H(k,N)$ distribution appears in Fig.
\ref{f_e_coli_network} which shows a uniform (age-independent)
distribution of regulated nodes over the genome, and this
uniformity is only expected for $\alpha=1$ corresponding to linear
growth in link numbers per node and quadratic growth in regulator
numbers.

\begin{figure}[htbp]
\centering
\includegraphics[width=\columnwidth,clip]{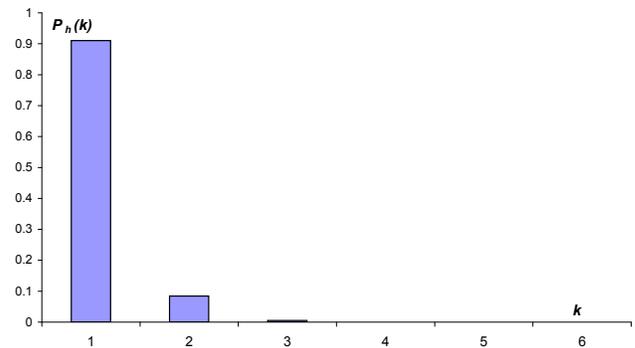}
\caption{\em The predicted proportions $P_h(k)$ of the regulated
operons of E. coli taking multiple regulatory inputs for a genome
of $N=2528$ operons. This distribution closely approximates that
observed for E. coli in Fig. 2(d) of Ref.
\protect\cite{Cherry_03_40} and of Fig. 5 of Ref.
\protect\cite{MandanBabu_03_1234}. \label{f_heads_per_reg_operon}}
\end{figure}

These predictions can be compared to observation. For the {\em E.
coli} network of size $N=2528$ operons or 4289 genes
\cite{Cherry_03_40}, the predicted proportion of regulated operons
receiving $k>0$ inputs is
\begin{equation}
    P_h(k) = \frac{H(k,N)}{1-H(0,N)},
\end{equation}
and is shown in Fig. \ref{f_heads_per_reg_operon}. Here, the
calculated distribution closely approximates the compact
exponential distribution observed for {\em E. coli} shown in Fig.
2(d) of Ref. \cite{Cherry_03_40} and of Fig. 5 of Ref.
\cite{MandanBabu_03_1234}, though it underestimates the numbers of
regulated operons with 4, 5, 6 and 7 inputs---essentially no
regulators are predicted to have 5 or more inputs for genomes of
size $N=2528$ operons. In addition, the average number of inbound
regulatory links per operon (for all operons) is $\langle
k\rangle=L/N=lN=0.19$, while the average number of inbound
regulatory links for regulated operons is $\langle
k_r\rangle=L/O_r\approx 1$.  A more accurate calculation using the
specific values for {\em E. coli} gives $\langle
k_r\rangle=L/O_r=1.10$, very close to the {\em E. coli} value of
1.5 or 1.6 noted in Refs.
\cite{Thieffry_98_43,Shen_Orr_02_64,MandanBabu_03_1234}.

\subsection{Scale-free distribution of regulator operons}

On entry into the genome, node $n_k$ sources on average $lk$
outbound regulatory links and this linear growth in link number
means that more recent nodes are more likely to be immediately
regulatory and more likely to be highly connected on genome entry.
However, node $n_k$ will also receive on average $lk$ inbound
regulatory links whose tails will be preferential attached to
existing regulators. The final distribution of link number with
age will depend on the rate at which earlier nodes under
preferential attachment can attract links relative to the linearly
increasing link numbers of later regulatory nodes.

On entry at time $k$, node $n_k$ receives $h_{kk}\approx lk$
inbound links from existing regulatory nodes in the set
$n_1,\dots,n_k$. As previously, we gain insight by considering the
general case where $t_{kk}\approx h_{kk}\approx lk^\alpha$ for
$\alpha\geq 0$ (though we continue to use the distribution
$P(j,k)$ of Eq. \ref{eq_P_j_k_dist} to determine both the number
of links $j$ prior to exponentiation and regulatory probability so
consequently the number of regulators continues to increase
quadratically according to Eq. \ref{eq_reg_density}). As a result,
the need to ensure that all regulatory links to node $n_k$ are
distinct requires that new link number $lk^\alpha$ be less than
the number of existing regulators $lk^2$ requires $\alpha\leq 2$.
The $h_{kk}$ new tails added with node $n_k$ are preferentially
attached to the existing regulatory nodes $n_j$ with probability
proportional to the number of existing regulatory links for that
node at time $k$, i.e. $t_{jk}$. Using the continuous
approximation
\cite{Barabasi_99_17,Barabasi_99_50,Dorogovtsev_01_056125}, the
rate of growth in outbound link number for node $n_j$ is then
approximately
\begin{equation}    \label{eq_continuum_tails}
  \frac{\partial t_{jk}}{\partial k} =
     h_{kk}\frac{t_{jk}}{\int_{0}^{k} t_{jk} \; dj}.
\end{equation}
The denominator here is a probability weighting to ensure
normalization and is the total number of outbound links for all
nodes. Following \cite{Dorogovtsev_02_10}, we can evaluate the
denominator using the identity
\begin{equation}
  \frac{\partial}{\partial k} \int_0^k t_{jk} \;dj =
     \int_0^k  \frac{\partial}{\partial k} t_{jk} \; dj + t_{kk}.
\end{equation}
This can be evaluated using Eq. \ref{eq_continuum_tails} noting
$t_{kk}\approx h_{kk}\approx lk^\alpha$ giving
\begin{equation}
  \frac{\partial}{\partial k} \int_0^k t_{jk} \;dj = 2lk^\alpha,
\end{equation}
which can be integrated determining the denominator of Eq.
\ref{eq_continuum_tails} to be
\begin{equation}
   \int_0^k t_{jk} \;dj = \frac{2l}{\alpha+1}k^{\alpha+1}.
\end{equation}
This is in agreement with Eq. \ref{eq_total_links}. Substituting
this value into Eq. \ref{eq_continuum_tails} gives
\begin{equation}
  \frac{\partial t_{jk}}{\partial k} =
     \frac{\alpha+1}{2} \frac{t_{jk}}{k}.
\end{equation}
Finally, this can be integrated with initial conditions
$t_{jj}\approx lj^\alpha$ at time $j$ and final conditions
$t_{jN}$ at time $N$ to give
\begin{equation}     \label{eq_tails_t_jn}
  t_{jN} = l  N^{\frac{\alpha+1}{2}} j^{\frac{\alpha-1}{2}}.
\end{equation}
Again we find that the respective choices $\alpha<1$ and
$\alpha>1$ lead to monotonically decreasing and increasing numbers
of links per node as a function of node number, while setting
$\alpha=1$ ensures the number of links per node is independent of
node number.  In this case, the preferential attachment of links
to earlier nodes does indeed act to cancel the initial bias in
link number towards later nodes. It is also apparent that when
$\alpha=1$, the limit $l\rightarrow 1$ implies all nodes possess
exactly $N$ links as expected for a fully connected regular
network.  (Preferential attachment cannot distort connectivity
numbers in this case as all nodes have an equal number of links.)
Additionally, in the limit $l\rightarrow 0$ we have $t_{jN}=0$ as
required for an entirely disconnected network.  The case
$\alpha=0$ duplicates results found for growing networks which add
a constant number of links with each new node subject to
preferential attachment \cite{Albert_02_47}.

As previously, it is straightforward to calculate the final
outbound link distribution in the case $\alpha=1$ using Eq.
\ref{eq_final_link_dist}.  This gives
\begin{eqnarray}
   T(k,N) &=& \frac{1}{N} \int_0^N dj \; \delta(k-lN)   \nonumber \\
   &=& \delta(k-lN).
\end{eqnarray}
Again, we find the expected compact distribution resulting when
all nodes possess the same average number of links. This raises
the question however, of how it is that a probabilistic
accelerating network subject to preferential attachment can end up
with all nodes possessing the same average number of links? The
answer lies in our use of two classes of distinguishable nodes,
regulators and non-regulators, which requires that we take into
account the known distribution of regulators with node number over
the genome. The average link number per node at node $n_j$ (Eq.
\ref{eq_tails_t_jn}) equates to the product of the average number
of link tails per regulator at node $n_j$, denoted $t_r(j,N)$, and
the average number of regulators per node at node $n_j$, denoted
$\rho(j)$.  This latter density is $\rho(j)=dR(j)/dj\approx lj$ by
Eq. \ref{eq_reg_density}, so by definition, we have
\begin{equation}
  t_{jN}= t_r(j,N) \rho(j),
\end{equation}
giving
\begin{equation}   \label{eq_t_ii_dist}
  t_r(j,N)= N^{\frac{\alpha+1}{2}} j^{\frac{\alpha-3}{2}}.
\end{equation}
Hence, for $\alpha<3$, the average number of links per regulator
is a decreasing function of node number $j$ as the growing number
of links added to recent nodes is insufficient to outweigh the
effects of preferential attachment which more rapidly increases
the number of links attached to early nodes.  In particular, for
$\alpha=1$ with the addition of a linearly increasing number of
links per node, the average number of regulatory links per
regulator scales inversely with node number $j$. In other words,
the density of regulators is very low at small node numbers $j$
while the very few regulatory nodes in this stretch of the genome
are heavily connected due to preferential attachment so as to
maintain the constant average of Eq. \ref{eq_tails_t_jn}. (See
Fig. \ref{f_e_coli_network}.)

\begin{figure}[htb]
\centering
\includegraphics[width=\columnwidth,clip]{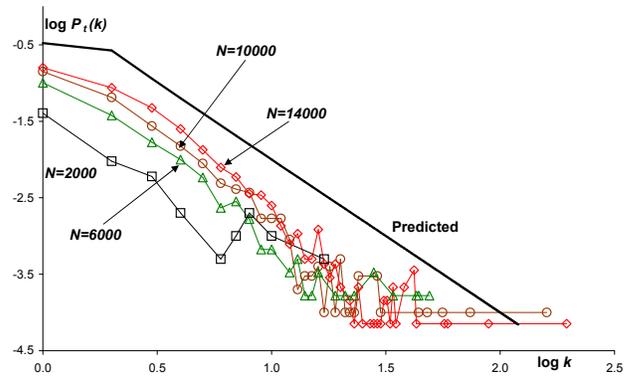}
\caption{\em  A simulation of the proportion of outbound links per
regulator $P_t(k)$ in networks of various sizes with linear growth
in the probable number of links per node preferentially attached
to regulatory nodes.  The log-log plot shows slopes of roughly
$-2$ in agreement with theoretical predictions (heavy solid line)
of a long-tailed scale free distribution $P_t(k)\propto k^{-2}$.
\label{f_simulation_tails}}
\end{figure}

\begin{figure}[htb]
\centering
\includegraphics[width=\columnwidth,clip]{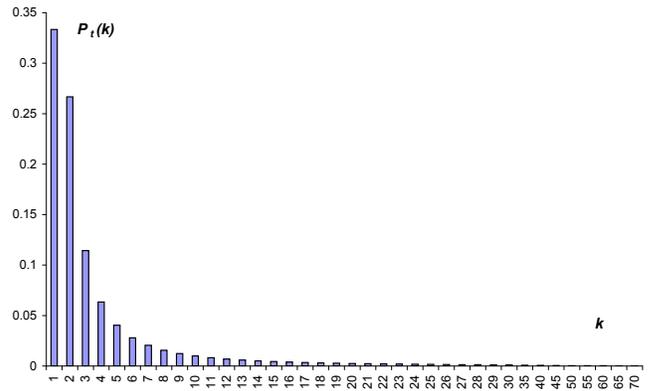}
\caption{\em The predicted proportion of regulatory operons
$P_t(k)$ regulating $k$ different operons for a simulated E. coli
genome with $N=2528$ operons.   As expected, most regulators
regulate only one other operon, though a small number of
regulators can regulate more than 40 operons. This distribution
closely approximates the observed proportions for E. coli in Fig.
2(c) of Ref. \protect\cite{Cherry_03_40} and Fig. 4 of Ref.
\protect\cite{MandanBabu_03_1234}, and predicts the probable
existence of about one E. coli regulator possessing link numbers
in each of the respective ranges between $[40,49]$ links, between
$[50,64]$ links, between $[65,94]$ links, between $[95,169]$
links, and between $[170,700]$ links, and so on.
\label{f_e_coli_long_tails}}
\end{figure}

The $t_r(j,N)$ distribution contains information about both node
connectivity and node age and so approximates genome statistics
(simulated or observed) when all of this information is available.
However, it is usually the case that node age information is
unavailable necessitating calculation of connectivity
distributions that are not conditioned on node age. This
effectively requires binning together all nodes irrespective of
their age to obtain a final link distribution.  In the case of
linearly growing number of links per node, $\alpha=1$, the delta
function of Eq. \ref{eq_final_link_dist} is resolved by the
equality $j=N/k$ giving the final distribution as
\begin{eqnarray}   \label{eq_k_minus_two}
   T(k,N) &=& -\frac{1}{N} \left( \frac{\partial j}{\partial k}\right)\nonumber \\
   &=& \frac{1}{k^2},
\end{eqnarray}
which, as required, is normalized as $\int_1^\infty T(k,N)=1$. The
expected proportion of regulators $P_t(k)$ possessing $k$ links is
then obtained by integrating the continuous distribution of Eq.
\ref{eq_k_minus_two} over appropriate ranges $[1,3/2]$ or
$[k-1/2,k+1/2]$ to obtain
\begin{equation}
    P_t(k) = \left\{
      \begin{array}{cc}
        \frac{1}{3} & k=1 \\
         &  \\
        \frac{4}{4k^2-1} & k>1. \\
      \end{array}
    \right.
\end{equation}
These theoretical predictions compare well to simulations of
networks of various sizes with linearly increasing numbers of
probable links per node and subject to preferential attachment.
Fig. \ref{f_simulation_tails} shows simulated outbound link
distributions which are long-tailed and scale free with
probabilities scaling roughly as $P_t(k)\propto k^{-2}$ for large
$k$. The $P_t(k)$ distribution shows a full one third of
regulators have only one link, while 60\% have two or fewer links,
and 71\% have three or fewer links. Fig. \ref{f_e_coli_long_tails}
shows the long-tailed distribution $P_t(k)$ expected for a
simulated {\em E. coli} network of $N=2528$ operons with
preferential attachment of links.  This figure shows marked
similarities to the long-tailed distribution of {\em E. coli}
shown in Fig. 2(c) of Ref. \cite{Cherry_03_40}. In particular, the
expected number of regulators with $k$ links is $P_t(k) R(N)$ with
the number of regulators $R(N)$ obtained from Eq.
\ref{eq_reg_density} (or from observation). For {\em E. coli},
this predicts the probable existence of about one regulator
possessing link numbers in each of the respective ranges between
$[40,49]$ links, between $[50,64]$ links, between $[65,94]$ links,
between $[95,169]$ links, and between $[170,700]$ links for
instance. (This approximates the connectivity of the global food
sensor CRP which regulates up to 197 genes directly
\cite{MandanBabu_03_1234}.) The average of the $P(k)$ distribution
(as well as the $t_r(j,N)$ distribution) is formally undefined as
long as the integration limits are taken to infinity. However, in
a network of $N$ nodes, a regulator can practically only regulate
a total of $N$ nodes, and this cutoff allows us estimate the
average connectivity per regulator (complementing previous
estimates following Eq. \ref{eq_l_assignment}). Using the cutoff
and approximating the summation via an integral, the average
connectivity per regulator in a network of $N$ nodes is
\begin{eqnarray}    \label{eq_pk_average}
  \langle k \rangle &=& \sum_{k=1}^{N} k P_t(K) \nonumber \\
    &=& \frac{1}{3} +
  \frac{1}{2} \ln \left( \frac{4N^2-1}{15} \right),
\end{eqnarray}
(or simply $\ln N$ using the continuous distribution of Eq.
\ref{eq_k_minus_two}.) The average number of links per regulator
for {\em E. coli} from Eq. \ref{eq_pk_average} is $\langle k
\rangle=7.51$ (or 7.83 using the simpler derivation), which again
compares well to the observed value of 5 in {\em E. coli}
\cite{Shen_Orr_02_64}.

\section{Inherent prokaryote size limits}
\label{sect_size_constraints}

\begin{figure}[htbp]
\centering
\includegraphics[width=\columnwidth,clip]{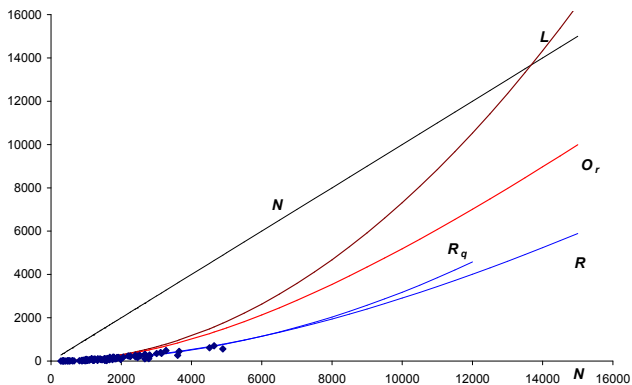}
\caption{\em  The quadratic growth in the number of regulatory
links $L$, and the asymptoting quadratic growth of regulatory
operons $R$ and of regulated operons $O_r$ in relation to the
total number of operons $N$. Actual numbers of regulators for 89
prokaryote genomes are shown (solid points), while the
non-asymptoting fitted quadratic curve $R_q$ is shown for
comparison. The observed maximum size of prokaryote genomes (of
order 10,000 genes or about 6,000 operons) lies near the
transition point between sparse and dense connectivity as an
increasing proportion of operons become linked into the regulatory
network. \label{f_prokaryote_model}}
\end{figure}

The accelerating nature of regulatory gene networks necessarily
means that these networks must exhibit a transition at some
critical network size either to a nonaccelerating architecture
permitting continued growth or must cease growth entirely, and we
now seek to predict the location of this transition point and
compare it to the evolutionary record.  We begin by examining an
overview of the accelerating genome model. Fig.
\ref{f_prokaryote_model} shows that linear growth in link numbers
per node ($\alpha=1$) allows a quadratic growth in the total
number of links (Eq. \ref{eq_link_No}) despite each of the number
of regulators (Eq. \ref{eq_reg_density}) and the number of
regulated nodes (Eq. \ref{eq_reg_operons}) asymptoting to some
fraction of $N$ after an initial period of quadratic growth. For
large genomes, almost all new nodes will be regulators and densely
connected into the existing network which will then multiply
regulate almost every node.

The transition from sparse to dense connectivity occurs as an
increasing proportion of operons become linked into the regulatory
network leading to the emergence of a single giant component of
fully connected nodes.  One way to highlight this transition is by
determining the proportion of transcription factor which control
downstream regulators as such linkages create the single giant
component.  The proportion of regulators controlling regulators is
\begin{eqnarray}    \label{eq_regulation_regulators}
   P_{rr}(N) &=& \frac{1}{R(N)} \sum_{k=1}^N
     \left[1-(1-l)^k \right] \frac{N}{k}\frac{R(N)}{N} \nonumber  \\
     &\approx & lN.
\end{eqnarray}
Here, the first fraction on the RHS normalizes the proportion in
terms of the number of regulators $R(N)$ (Eq.
\ref{eq_reg_density}), the first term in the summation is the
probability that node $n_k$ is a regulator, the second term is the
average number of regulatory outbound links for this regulatory
node $t_r(k,N)$ at network size $N$ (Eq. \ref{eq_t_ii_dist} with
$\alpha=1$), and the third term approximates the probability that
these nodes link to one of the existing regulators under random
attachment.  (If the very first and very last terms are dropped,
the remaining summation over all nodes of the probability that
$n_k$ is regulatory with the stated number of links equates to the
total number of links in the network $L\approx lN^2$. This is the
more accurate version of the calculation leading to Eq.
\ref{eq_t_ii_dist}.) Hence, the proportion of regulators which
control transcription factors scales linearly with network size
and equals 15\% for an $N=2000$ network, 29\% for $N=4000$, 44\%
for $N=6000$, 59\% for $N=8000$, 73\% for $N=10000$, and 88\% for
$N=12000$ operons (after which the approximations made break
down). Naturally, when most regulators themselves control other
regulators, then the entire regulatory network will consist of a
single giant component. These ratios compare reasonably well with
those observed in {\em E. coli} where Ref.
\cite{MandanBabu_03_1234} noted that of 121 transcription factors
for which one or more regulatory genes are known, 38 factors or
31.4\% regulate other transcription factors. The approximate
second line of Eq. \ref{eq_regulation_regulators} with $N=2528$
for {\em E. coli} determines this proportion as $P_{rr}=18.5\%$
while the more accurate top line gives the proportion of
regulators which control transcription factors as $P_{rr}=17.7\%$,
giving a reasonable match between prediction and observation.

As the proportion of regulators of transcription factors rises,
the probable length of regulatory cascades will increase. In fact,
the proportion of regulators taking part in a regulatory cascade
of length $n\geq 1$ is
\begin{equation}
   p_n = (1-P_{rr}) P_{rr}^{n-1}.
\end{equation}
This equation can be obtained from a tree of all binary pathways
which at each branching point either terminate with probability
$(1-P_{rr})$ or cascade with probability $P_{rr}$.  As such, the
probable cascade length is negligible when the proportion of
regulators controlling regulators is small $P_{rr}\ll 1$ but can
become large as $P_{rr}$ itself increases.  As $P_{rr}$ is indeed
large for networks of size $N>6000$, this again suggests that long
cascades of regulatory interactions will lead to the coalescing of
a single giant component in this regime. Again, the calculated
lengths of regulatory cascades can be compared to those in {\em E.
coli} where the number of cascades of regulated transcription
factors observed in a particular set of regulatory interactions
was 23 two-level cascades or 37.7\%, 32 three-level cascades or
52.5\%, and 6 four-level cascades or 9.8\%
\cite{MandanBabu_03_1234}.  As one-level or autoregulatory
interactions are not included in this observation, the predicted
proportions for {\em E. coli} are $\bar{p}_n=p_n/(1-p_1)$ with
$P_{rr}=17.7\%$ giving 82\% two-level cascades, 15\% three-level
cascades, 3\% four-level cascades, 1\% five-level cascades, and so
on.  It is seen that the theoretical predictions overestimate the
proportion of two-level cascades and underestimate the number of
three-level and higher cascades probably because of selection
pressures not included in the model. Lastly, we note that the
number of cycles involving closed regulatory loops of size greater
than one (i.e. involving more than autoregulation) in the examined
portion of the {\em E. coli} regulatory network is zero reflecting
that feedback loops in these organisms are carried out at the
post-transcriptional level involving metabolites such as appear in
the {\em lac} operon
\cite{Thieffry_98_43,Shen_Orr_02_64,MandanBabu_03_1234}.

We note that our model is entirely unable to explain the high
proportion of autoregulation observed in {\em E. coli} with
various estimates that 28.1\% \cite{Rosenfeld_02_785}, 50\%
\cite{Shen_Orr_02_64} and 46.9\% \cite{Thieffry_98_43} of
regulators are autoregulatory.  The predicted proportion of
autoregulators is approximated by replacing the very last fraction
($R/N$) in Eq. \ref{eq_regulation_regulators} by the term $1/N$
giving the probability that a self-directed link is formed,
leading to the expected autoregulatory proportion $\approx
2/N\approx 0.08\%$ for {\em E. coli}.  This failure likely
reflects the action of selection processes promoting spatial
rearrangements of entire regulons on the genome and the internal
shuffling of genes and promotor units. Such reorganizations of
duplicated gene regions (presumably shuffling genes and promotor
regions) have been common in {\em E. coli} allowing for instance,
spatial regulatory motifs whereby the promotors of colocated
(overlapping) and often co-functional operons transcribed in
opposing directions can interfere \cite{Warren_03_0310029}.

The transition point from sparse to dense connectivity can be
roughly located using the continuum approximation
\cite{Barabasi_99_17,Barabasi_99_50,Dorogovtsev_01_056125}. These
methods have not previously been used for this purpose (to our
knowledge) and we first validate their use by deriving the known
result that non-growing random graphs of $N$ nodes connected by an
increasing number of $L$ undirected links undergo a phase
transition from sparse to dense connectivity when $L=N/2$
\cite{Erdos_60_17}. As the number of links $L$ grows, the $N$
nodes are interlinked into firstly separate islands of size $s_i$
nodes for $i=1,2,\dots$ which eventually link up to form a giant
component designated $s_1$ containing essentially all nodes
$s_1\approx N$. The largest component grows whenever a newly added
link has either its head or tail in island $s_1$ (with probability
$s_1/N$) and the other outside it (with probability $(N-s_1)/N$)
leading to a size increment equal to the average size of the
external islands ($\langle s_{j\neq 1} \rangle$), giving
\begin{equation}
  \frac{ds_1}{dL} = \left[2 \frac{s_1}{N} \frac{(N-s_1)}{N}\right]
       \langle s_{j\neq 1} \rangle.
\end{equation}
Numerical or analytic integration of this equation with initial
conditions $s_1=2$ when $L=1$ and assuming the average size of
external smaller islands is $s_1/2$ shows the largest island
saturating the entire network when $L=N/2$ as expected. (This
simple approach is indicative only and is quite sensitive to for
instance, the assumed average size of external islands.)

This result suggests the following transition point in directed
regulatory gene networks. Each undirected (i.e. bidirectional)
link in random graph theory is equivalent to two directed links
allowing bidirectional traffic between any two nodes, suggesting a
transition point in directed graphs at roughly $L=N$. This
analysis suggests that the largest component is expected to
saturate the entire network when link number $L\approx N$ or
$N=1/l=13,677$ (see Fig. \ref{f_prokaryote_model}). In turn, this
suggests that for $N<13,677$ a typical network likely consists of
isolated trees, while if $N>13,677$ the network likely consists of
a single giant cluster where almost every node is connected to all
others via intermediate links. When the link number is very large,
$N\gg 13,677$, then the network becomes regularly connected
\cite{Albert_02_47}. As prokaryote regulatory networks likely
consist of functionally distinct regulated modules
\cite{Thieffry_98_43,Hartwell_99_c4}, it is unlikely that
prokaryotic gene networks can successfully operate in the fully
connected regime suggesting that prokaryote genome sizes are size
constrained $N\leq 13,677$.  In fact, the previously noted absence
of regulatory cycles in {\em E. coli}
\cite{Thieffry_98_43,Shen_Orr_02_64,MandanBabu_03_1234} likely
reflects the evolutionary importance of maintaining disjoint and
non-interfering regulatory units.

These results of random graph theory are suggestive only, and we
now turn to consider the size of the largest connected island in
prokaryote gene networks featuring directed links whose tails are
preferentially attached to regulators and whose heads are randomly
distributed over all existing nodes.  A further difference is that
prokaryote regulatory networks are themselves growing with each
added node accompanied by a probabilistic number of links. In
addition, we define an island to consist of all nodes which are
linked regardless of the orientation of all links and so
effectively treat links as being undirected. This is because a
regulator can potentially perturb every node downstream to it
including those nodes downstream of other regulators and so can
modify the regulatory effects of other regulators---essentially,
if the downstream effects of different regulators eventually
intersect, we count these regulators in the same island. (Other
definitions of islands could be used.)

\begin{figure}[htbp]
\centering
\includegraphics[width=\columnwidth,clip]{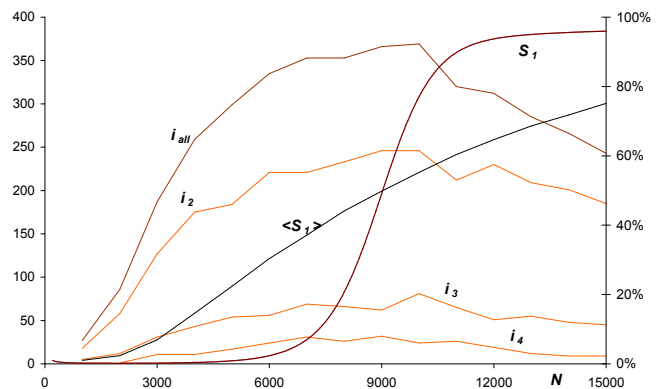}
\caption{\em The total number of discrete disconnected islands
$i_{\rm all}$, the number of islands with respectively two
($i_2$), three ($i_3$) and four ($i_4$) members (left hand axis),
and the simulated ($\langle s_1\rangle$) and predicted ($s_1$)
size of the largest island measured as a proportion of nodes for
various genome sizes (right hand axis). The simulations show the
largest island contains $\langle s_1\rangle=50\%$ of all nodes at
a critical network size of $N_c=9,029$ nodes. The input parameters
of the predicted curve $s_1$ are set so $s_1=\langle s_1\rangle$
at this point.}\label{f_largest_island}
\end{figure}

The dominant (but not sole) mechanism by which island $s_1$ can
grow is for the newly added node $n_k$ to either (a) be a
regulator (with probability $[1-(1-l)^k]$) and establish an
outbound regulatory link to some existing node in $s_1$ (with
probability $s_1/k$) while at the same time accepting a regulatory
link (with probability $[1-(1-l)^k]$) from a node in a different
island $s_{j\neq 1}$ (with probability $(k-s_1)/k$), or (b) accept
an inbound regulatory link (with probability $[1-(1-l)^k]$) from a
regulator in island $s_1$ (with probability $s_1/k$) while
establishing a regulatory link (with probability $[1-(1-l)^k]$) to
some node in a different island $s_{j\neq 1}$ (with probability
$(k-s_1)/k$). (Here, we assume that regulators are uniformly
distributed over islands and the number of links within an island
scales with the size of the island to crudely model preferential
attachment.) The result is that island $s_1$ grows by the size of
the second island assumed to be $s_{j\neq 1}$. Altogether, the
rate of growth in the size of island $s_1$ is then
\begin{equation}
  \frac{ds_1}{dk} =
     2 \left[ 1 - (1-l)^k \right]^2
     \frac{s_1 [k-s_1]}{k^2} \langle s_{j\neq 1} \rangle.
\end{equation}
For initial conditions, we assume that a first link appears when
the genome has $(pg_0^2)^{-1/2}=177$ nodes ($s_1(177)=2$).
Simulations show that sufficient small islands are created to
ensure $\langle s_{j\neq 1} \rangle$ remains roughly constant and
equal to $\langle s_{j\neq 1}\rangle=2.72$, though matching the
simulated and predicted curves at the 50\% point requires setting
$\langle s_{j\neq 1}\rangle=30$. This is reasonable given the
approximations made. Fig. \ref{f_largest_island} shows the size of
the largest island $s_1$ as a proportion of all nodes. A single
giant component is expected to form at a critical genome size of
$N_c=9,029$ operons defined as the point where the simulated
proportion of nodes in the giant component is 50\%. (Choosing a
parameter setting of 40\% would also be justifiable and would lead
to an exact match between predicted and observed maxima.) Unlike
random graph theory, this critical point applies to all growing
genomes as it is determined by the value of the link formation
probability $l$. Genomes of smaller size than this critical value
$N<N_c$ are expected to be sparsely connected so the network
consists of multiple discrete connected islands (as in {\em E.
coli} \cite{Thieffry_98_43}), while genomes of larger size $N>N_c$
are expected to be densely connected into a single giant component
where every regulator eventually perturbs the downstream effects
of every other regulator.

Simulations of example genomes of various sizes spanning this
critical network size confirm the adequacy of the continuum
treatment.  Fig. \ref{f_largest_island} shows the number of all
discrete islands as well as the number of islands containing two,
three and four components.  In the vicinity of the critical genome
size $N_c=9,029$, the number of discrete interconnected islands
begins to decline as the growing number of links connects more and
more islands into the single giant component.  The size of the
simulated giant component as a proportion of genome size is also
shown. This figure suggests that the {\em E. coli} genome of
$N=2528$ operons should possess a giant component containing about
5\% of all nodes (about 100 nodes) which can be compared with the
observation that about 70\% or 300 operons of the examined
regulatory and regulated operons (but not including unregulated
and nonregulatory operons) could be loosely grouped into 3-6
``dense overlapping regulons" or DORS while the remaining operons
appeared as disjoint systems with most containing 1-3 operons but
some containing up to 25 operons \cite{Shen_Orr_02_64}.

The critical network size of $N_c=9,029$ operons or about
$N_g=15,349$ genes corresponds to the point where growing
regulatory networks exploiting accelerating links can no longer
maintain discrete functional units, islands, of interconnected
nodes. Larger genomes are densely connected into a single giant
component where, eventually, any regulator can perturb the
downstream effects of every other node so for instance, it is
unlikely that the discrete network motifs found in the {\em E.
coli} regulatory network \cite{Shen_Orr_02_64} can survive in this
regime. This massive increase in perturbative effects immeasurably
increases the difficulty of the evolutionary search process,
leading to an expectation that the rate of evolutionary change
will drastically slow when growing genome sizes reach criticality
$N\approx N_c$. From a biological point of view, it is relatively
easy to understand why the critical network size $N_c$ acts as an
upper size limit.  The accelerating nature of the prokaryote
regulation network means that larger networks can add new nodes
only be integrating an increasing number of links to gain
evolutionary benefits. Of course, the probability of finding $lN$
beneficial links is a rapidly decreasing function of $N$.  It is
relatively easy to find a beneficial regulator making only of
order one link to existing genes (only billions of trials are
needed say), but much harder when the regulator is making an
average of five links with existing genes (many trillions of
trails are needed). Essentially, the more links that must be
beneficially integrated, the longer the evolutionary search task
and the slower the rate of evolutionary change.

\begin{figure}[htbp]
\centering
\includegraphics[width=\columnwidth,clip]{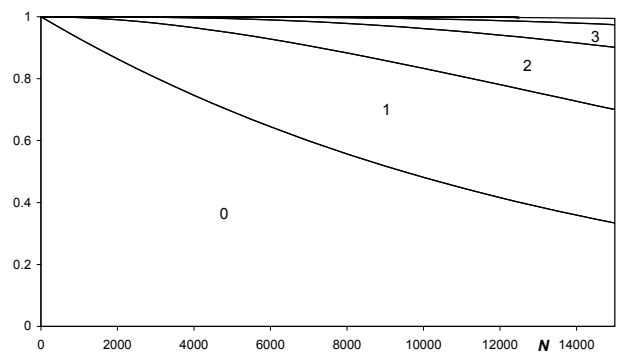}
\caption{\em The predicted proportion $P_o(k,N)$ of operons with
$0, 1, 2, \dots$ regulatory inputs as a function of network size.
Small networks mainly possess unregulated operons, while networks
of large size have a significantly reduced number of unregulated
operons with many operons taking large numbers of regulatory
inputs. \label{f_prob_reg_operon_history}}
\end{figure}

Many other statistical measures suggest that the regulatory
mechanisms optimized to perform in a sparsely connected network
will not necessarily operate in a densely connected
network---evolution cannot foresee later needs.  In particular,
the proportion of operons $n_j$ which are regulated by $k$ inputs
is, using Eq. \ref{eq_h_kNo_final}, given by
\begin{equation}
  P_o(k,N)= \frac{1}{N}\sum_{j=1}^N H(k,N) \;=\;
      {N \choose k} l^k (1-l)^{N-k}.
\end{equation}
This distribution increases with increasing network size and is
shown in Fig. \ref{f_prob_reg_operon_history} making it clear that
small networks mainly possess operons which are either entirely
unregulated or regulated by only one or a few regulators. In
contrast, large networks ($N>N_c$) have only a small proportion of
operons which are unregulated while the majority of operons take
between one or more regulatory inputs. It is a more difficult
evolutionary task to integrate many inputs to achieve a
beneficially regulated output again suggesting that prokaryote
regulatory networks featuring accelerating growth in link number
are size limited due to their regulatory architecture.

\begin{figure}[htbp]
\centering
\includegraphics[width=\columnwidth,clip]{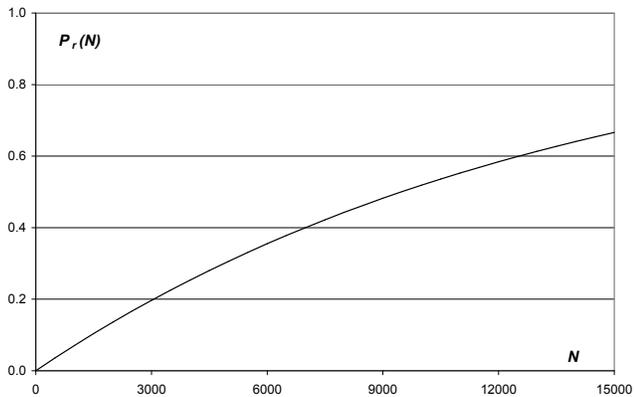}
\caption{\em  The rapidly increasing probability $P_r(N)$ that the
most recently added node $n_{N}$ in a network of size $N$ nodes is
immediately regulatory on its appearance in the genome. For
network sizes greater than about $N_c=9,029$ operons, the
probability that all new nodes are immediately regulatory exceeds
about 50\%. \label{f_prob_regulator_No}}
\end{figure}

Another way to suggest the strict size limits imposed by the
accelerating growth of regulatory links is to consider the
probability that the most recently added node $n_{N}$ in a network
of size $N$ immediately becomes regulatory.  Using Eq.
\ref{eq_P_j_k_dist}, node $n_{N}$ is a regulator with probability
\begin{equation}
  P_r(N)=\sum_{i=1}^N P(i,N) \;=\;  1 - (1-l)^N.
\end{equation}
This probability tends to unity as network size increases, and in
particular, surpasses about 50\% when networks consist of $N_c$
operons---see Fig. \ref{f_prob_regulator_No}.  At about this
stage, large networks cannot add a new node without it having a
significant probability of modifying the dynamics of existing
nodes. This immeasurably increases the difficulty of the
evolutionary task and again suggests a maximum size limit to
prokaryote gene regulatory networks.

If the accelerating regulatory networks of prokaryotes were able
to operate in the densely connected regime, the evolutionary
record might be expected to show prokaryotes of arbitrarily large
genome size with a transition in connectivity statistics at some
critical genome size of about $N_c\approx 9,029$.  Conversely,
should these regulatory networks be unable to operate in the
densely connected regime, then the evolutionary record should show
a maximum size limit to prokaryote genome sizes of about
$N_c\approx 9,029$ operons or about $N_g=15,349$ genes, close to
the observed upper limit.

\section{Conclusion}

In this paper, we generalize models of accelerating networks by
including probabilistic links to allow arbitrarily rapid
acceleration rates leading to structural transitions in growing
networks sometimes severe enough to strictly constrain network
size. These structural transitions from sparse to dense
connectivity are made more difficult by any additional steric or
logical limitations on combinatoric control at any given promotor.
Such transitions are in sharp contrast to the stationary
statistics and unbounded growth potential of non-accelerating
scale free and exponential networks. These probabilistic
accelerating networks were applied to model prokaryote regulatory
networks which exploit a quadratic growth in the number of
regulators and regulatory links with genome size as established
via comparative genomics programs.  Our models predict a maximum
genome size of $N_c\approx 9,029$ operons or about $N_g=15,349$
genes for prokaryotes, closely approximating the observed maximum.
We further validated our model by making a detailed comparison of
predicted and observed results for {\em E. coli}, and achieved
satisfactory matches for respectively, the number of observed
regulators, an average promotor binding site length of about 7,
the long tailed distribution of outgoing regulatory links with an
average of between 2.12 and 7.51 (compared to 5), the exponential
distribution of incoming regulatory links with an average of
around 1.10 (compared to 1.5), the proportion of regulators
controlling regulators of around 17.7\% (compared to 31.4\%), and
the probable length of regulatory cascades and the absence of
regulatory loops.  Our approach is unable to explain the high
proportion of autoregulation observed in {\em E. coli}
\cite{Shen_Orr_02_64} and this failure likely points to selection
for genome reorganizations leading to spatial arrangements of
operons allowing joint regulation \cite{Warren_03_0310029} which
is not included in this model. Further, this approach does not
include selection pressures ensuring that similarly regulated
islands or modules share common functionality
\cite{Shen_Orr_02_64}, or other regulatory mechanisms influencing
both the transcription and translation of transcription factors
including micro-RNAs and other chemical mechanisms and mediators
(see for instance \cite{Vogel_03_6435}).

However, the many successes of the accelerating network model of
prokaryote regulatory networks are meaningless if similar results
can be achieved via non-accelerating network models.  In later
work, we will show that the two simplest non-accelerating network
models fail to explain either the observed quadratic growth of
regulator number with genome size or the detailed statistics
pertaining to the {\em E. coli} genome \cite{Gagen_0312022}. In
addition, the simplifying assumption adopted here that gene
duplications ensure that operons become regulatory only on entry
to the genome will be dropped in later work. This will develop a
more realistic model including separate physical processes for
transcription factor binding to DNA and for establishing
regulatory links with regulated operons where links can form at
any time.

This work has wider significance due to the still common
presumption in molecular biology that ``What was true for {\em E.
coli} would also be true for the elephant" capturing the notion
that the mechanisms operating in prokaryotes are exactly identical
to those operating in complex multicellular eukaryotes. In this
picture, eukaryotes are merely enlarged prokaryotes.  The results
of this paper indicate that this is not possible---the
accelerating nature of regulatory networks necessarily implies
that eukaryotes cannot be scaled up prokaryotes and that the
(likely) accelerating regulatory networks of eukaryotes must be
exploiting novel regulatory mechanisms.  The successful modelling
of these mechanisms will likely require incorporating
computationally complex technologies
\cite{Mattick_01_1611,Mattick_01_986,Mattick_03_930} into an
accelerating network model, and this also will be addressed in
later work.


\end{document}